# Pressure-Stabilized $MnSb_2$ with Complex Incommensurate Magnetic Order

Mingyu Xu[1,2,3], Matt Boswell[1,4], Qing-Ping Ding[2], Cheng Peng[1], Aashish Sapkota[2,3], Qiang Zhang[4], Danielle Yahne[4], Sergey. L. Bud'ko[2, 3], Yuji Furukawa[2,3], Paul. C. Canfield[2,3], Raquel A. Ribeiro[2,3], Weiwei Xie[1*]

[1]Department of Chemistry, Michigan State University, East Lansing, Michigan 48824, USA

[2]Ames National Laboratory, Iowa State University, Ames, Iowa 50011, USA

[3]Department of Physics and Astronomy, Iowa State University, Ames, Iowa 50011, USA

[4]Neutron Scattering Division, Oak Ridge National Laboratory, Oak Ridge, Tennessee 37831, USA

Corresponding Authors: Weiwei Xie (xieweiwe@msu.edu)

***Abstract***

Marcasite-type compounds have been proposed as promising hosts of exotic magnetic quantum states, yet experimental realizations in stoichiometric, disorder-free systems remain limited. Here, we report the high-pressure stabilization and magnetic characterization of $MnSb_2$, a marcasite-type compound that is thermodynamically metastable under ambient pressure. Single crystals were synthesized using a cubic multi-anvil press, and powder and single-crystal X-ray diffraction confirm the orthorhombic *Pnnm* structure. These crystals are stable at ambient pressure for a long time up to between 450-500 K. Heat-capacity measurements reveal phase transitions at approximately 220 K and 118 K. Neutron diffraction uncovers an unconventional magnetic state below 220 K. Magnetic powder neutron diffraction refinements reveal possible multiple magnetic configurations that provide comparably acceptable fits to the experimental data. While most solutions are consistent with a spin-density-wave (SDW) description, helical models systematically yield inferior agreement factors. Across a broad range of models, the Mn ordered moment reaches a maximum value of approximately 2 $\mu_B$ and remains predominantly collinear, with minimal canting along the *c*-axis. At 200 K, the magnetic propagation vector is $q$ = (0, 0.3975, 0.3783); upon cooling, the b component increases toward 0.5, reflecting a temperature-dependent evolution of the modulation. The need for modification of the magnetic model between high and low temperatures further highlights the complex and strongly temperature-dependent nature of the magnetic order in this system. These results establish $MnSb_2$ as a pressure-stabilized marcasite magnet with a tunable, complex magnetic state and a compelling stoichiometric platform for exploring unconventional magnetic behavior, including potential altermagnetism.

## Introduction

When magnetism intertwines with band topology in quantum materials, it can, in some cases, give rise to a recently identified class of magnetic systems characterized by collinear spin arrangements with zero net magnetization, exhibiting time-reversal-symmetry-breaking responses[1–8] and spin-split electronic band structures [9–16]. These properties, referred to as altermagnetism, combine zero-net-moment magnetic order with symmetry-allowed spin splitting relevant for spintronic applications and have motivated intense theoretical and experimental interest in identifying chemically clean material platforms that realize this magnetic phase[8,17–31]. Marcasite-type compounds, in particular, have emerged as promising candidates due to their low crystal symmetry and symmetry-allowed spin splitting in antiferromagnetic states. $FeSb_2$ is a narrow-gap, strongly correlated semiconductor[32–36] that has been theoretically predicted to host a collinear altermagnetic state upon Co substitution or hole doping[37]. Experimentally, however, $FeSb_2$ exhibits no long-range magnetic order between 1.8 and 300 K. Stabilizing magnetic order in $FeSb_2$ therefore requires deliberate tuning of its electronic structure. While chemical substitution and external pressure are widely used to modify band filling and band width, substitution often introduces chemical disorder that complicates interpretation, and pressure-dependent magnetic measurements are experimentally nontrivial, especially for antiferromagnetism. This all motivates the search for chemically clean routes to stabilize magnetic order in the $FeSb_2$ structure class.

A promising route to this goal is provided by $MnSb_2$, a high-pressure marcasite-type compound that shares the same orthorhombic structure as $FeSb_2$ but incorporates Mn with a larger, and often more robust, local magnetic moment and intrinsic hole doping[38](relative to $FeSb_2$). Whereas $MnSb_2$ does not form at ambient pressure, it can be synthesized under high-pressure conditions and quenched into a metastable form at ambient pressure[39]. Across the marcasite family, $CrSb_2$ exhibits robust antiferromagnetic order with a Néel temperature near 273 K[40], whereas $FeSb_2$ remains nonmagnetic, suggesting that magnetic order emerges through band filling by hole doping. By analogy, $MnSb_2$ represents a chemically ordered, stoichiometric platform in which unconventional states may be realized.

## Experimental Methods

**High-Pressure Synthesis:** $MnSb_2$ was synthesized using a two-step method, which is different from the reference[39](lower pressure and temperature used in this experiment), involving (i) the preparation of a finely mixed Mn–Sb precursor and (ii) a subsequent high-pressure, high-temperature reaction. In the first step, manganese metal (CZ-1-110) and antimony shot (99.999%) were combined in a 1:2 molar ratio and loaded into a 1.7 mL fritted alumina Canfield Crucible Set (CCS, LSP Industrial Ceramics, Inc.).[41] The crucible was sealed in a silica ampoule under an argon atmosphere (~1/3 atm). Silica wool was packed around the CCS to stabilize the crucible during centrifugation and to contain any escaped liquid. The ampoule was heated in a box furnace to 760 °C over two hours, held at this temperature for 10 hours, and then rapidly centrifuged to separate the molten and non-molten phases. The solid residue retained on the frit disc appeared as a black solid, likely a minor oxide crust, while the solidified, decanted melt was collected, ground into a fine powder, and pressed into a pellet using a die press. In the second step, the pellet was loaded into a boron nitride (BN) crucible (7 mm length, 5.7 mm inner diameter), with any remaining void space filled with BN powder. The assembly was subjected to 3.3 GPa pressure at room temperature using a Rockland Research cubic multi-anvil press (~20mm anvil size) and then heated to 490 °C. This temperature was selected as the maximum before the Mn-Sb mix begins to melt, determined using the sample current (power) as a function of temperature curve, as shown in **Fig. S7*c***. The reason to prevent the mixture from melting during the experiment is to avoid a large amount of $Mn_{1.1}Sb$ impurity from incongruent melting, as shown in **Figs. S7*b*** and ***d***. After maintaining this temperature for 24 hours, the sample was quenched to room temperature before slowly releasing the pressure. This two-step process yields a high-quality $MnSb_2$ product, with more than half of the final material consisting of sub-millimeter-sized single crystals that can be easily separated from the surrounding polycrystalline matrix.

**Single Crystal X-ray diffraction:** To determine the crystal structure of the obtained single crystal, the sample with dimensions 0.218 × 0.158 × 0.138 $mm^3$ was picked up, mounted on a nylon loop with Paratone oil, and measured using an XtalLAB Synergy, Dualflex, Hypix single crystal X-ray diffractometer with an Oxford Cryosystems 800 low-temperature device. Data were collected using ω scans with Mo Kα radiation (λ = 0.71073 Å). The total number of runs and images was based on the strategy calculation from the program CrysAlisPro 1.171.43.92a (Rigaku OD, 2023). Data reduction was performed with correction for Lorentz polarization. The integration of the data using an orthorhombic unit cell yielded a total of 3540 reflections to a maximum θ angle of 40.13° (0.55 Å resolution), of which 478 were independent (average redundancy 7.406, completeness = 98.8%, Rint = 5.61%). A numerical absorption correction was applied based on Gaussian integration over a multifaceted crystal model[42]. Empirical absorption correction used spherical harmonics, implemented in the SCALE3 ABSPACK scaling algorithm.[43] The structure was solved and refined using the SHELXTL Software Package[44,45].

**Powder X-ray diffraction:** To examine the phase information, the powder X-ray diffraction (PXRD) analysis was performed subsequent to the synthesis process. The crystals were ground using an agate mortar and pestle to achieve a homogeneous powder. This powdered sample was then uniformly distributed on a single crystalline silicon sample holder, designed for zero background measurements, with a minimal application of vacuum grease to secure the powder in place. The PXRD data acquisition spanned a 2θ range from 15° to 80°, utilizing incremental steps of 0.01° and a fixed dwell time of 3 seconds per step. These measurements were conducted using a Rigaku MiniFlex II powder diffractometer, employing Bragg-Brentano geometry coupled with Cu Kα radiation (λ = 1.5406 Å). The refinement of the powder X-ray data was executed using the GSAS-II software suite[46].

**Physical Properties Measurements:** Temperature-, magnetic-field-dependent DC magnetization data and resistance measurements were collected using Quantum Design (QD), Magnetic Property Measurement Systems (MPMS and MPMS3), and Physical Property Measurement Systems (PPMS). During the measurements of single crystals, the field is along a random direction of the crystals due to the poorly defined facets and the small size of the crystal. The samples are placed between two collapsed plastic straws, with the third uncollapsed straw providing support as a sheath on the outside or a quartz sample holder. Samples were fixed on the straw or quartz sample holder with GE-7031-varnish. AC electrical resistance measurements were performed in a standard four-contact geometry using the ACT option of the PPMS, with a 3-mA current and a frequency of 17 Hz. 50μm diameter Pt wires were bonded to the samples with silver paint (DuPont 4929N) with contact resistance values of about 2-3 Ohms. Temperature-dependent specific heat measurements on the $MnSb_2$ in a mass of about 6 mg polycrystalline sample were carried out using a Quantum Design, Physical Property Measurement System (PPMS DynaCool) in the temperature range of 200-280 K ( H grease), 150 K- 280 K (N grease with addenda measurement points that are the same as the sample measurement), 107 K-127 K( N grease), and 4 K- 200 K (N grease).

**Neutron Powder Diffraction:** To determine whether magnetic ordering exists, neutron powder diffraction (NPD) measurements were performed using a time-of-flight powder diffractometer, POWGEN, at the Spallation Neutron Source at Oak Ridge National Laboratory. Approximately 2.6 g of samples were prepared after grinding several single crystals to a fine powder and passing them through a 45μm mesh sieve. Diffraction patterns were collected at 308 K for 2 hours 36 minutes using a neutron beam with a center wavelength of 1.5. Rietveld refinements using GSAS-II software[46] and Fullprof suite[47] were performed to determine the crystal and magnetic structures, respectively. Determination of the magnetic structure was performed on HB-2A POWDER beamline at the High-Flux Isotope Reactor (HFIR) located at Oak Ridge National Lab (ORNL). Roughly 1 g of $MnSb_2$ was loaded into an aluminum can backfilled with He. Diffraction patterns were measured with a vertically focused Ge monochromator to select the 2.41 Å wavelength while a collimation of open-open-12 was used. Magnetic structure analysis was performed with SARAh Representation Analysis and SARAh Refine and refined with the Fullprof suite software[48].

**Nuclear Magnetic Resonance:** Nuclear magnetic resonance (NMR) measurements of $^{55}$Mn (nuclear spin $I$ = 5/2, gyromagnetic ratio $\gamma_N/2\pi$ = 10.50 MHz/T), $^{121}$Sb ($I$ = 5/2, $\gamma_N/2\pi$ =10.189 MHz/T), and $^{123}$Sb ($I$ = 7/2, $\gamma_N/2\pi$ =5.517 MHz/T) nuclei were conducted using a laboratory-built phase-coherent spin-echo pulse spectrometer on polycrystalline powder samples. The NMR spectrum was obtained under magnetic fields by sweeping the magnetic field at a fixed NMR frequency of 67 MHz, while the zero magnetic field NMR spectrum was measured by plotting spin-echo intensity as a function of NMR frequency.

**DFT Calculation:** Density Functional Theory (DFT) calculations were carried out using version 7.3.1 of the Quantum ESPRESSO package to evaluate the total energies of $MnSb_2$ in non-magnetic, ferromagnetic, and antiferromagnetic configurations.[49] The calculations employed ultrasoft pseudopotentials and the Perdew-Burke-Ernzerhof (PBE) exchange-correlation functional within the generalized gradient approximation (GGA).[50–52] A plane-wave kinetic energy cutoff of 300 Ry was used, with a charge density cutoff set to 3600 Ry (12 times the wavefunction cutoff). Brillouin zone integration was performed using a 7×7×13 Monkhorst-Pack k-point grid.[53] Total energy convergence was ensured with a threshold of less than 1 meV per atom. The electronic self-consistency was achieved using the Davidson diagonalization algorithm, with a convergence criterion of $10^{-9}$ Ry.[54]

High-symmetrical *k*-point paths for band structure calculations were generated using the Spglib library.[55,56]

## Results and Discussion

$MnSb_2$ was synthesized at 3.3 GPa pressure and 490 °C using a Rockland Research cubic multi-anvil press. After maintaining this temperature for 24 hours, a high-quality $MnSb_2$ product, with more than half of the final material consisting of sub-millimeter-sized $MnSb_2$ single crystals that can be easily separated mechanically from the surrounding $MnSb_2$ polycrystalline matrix, was obtained. (See the Experimental Method section for further details.) The single-crystal X-ray diffraction (SCXRD) analysis of $MnSb_2$, summarized in **Tables S1** and **S2**, illustrates the crystal structure of $MnSb_2$ shown in the right inset of **Fig. 1**, which adopts the orthorhombic *Pnnm* space group and features edge-sharing $MnSb_6$ octahedra. The unit cell contains two crystallographically distinct atomic sites: Mn atoms occupy the 2*a* Wyckoff position, while Sb atoms reside at the 4*g* positions. Vacancies and site mixing were considered during the refinement, but no structural disorder or clear vacancies were detected.

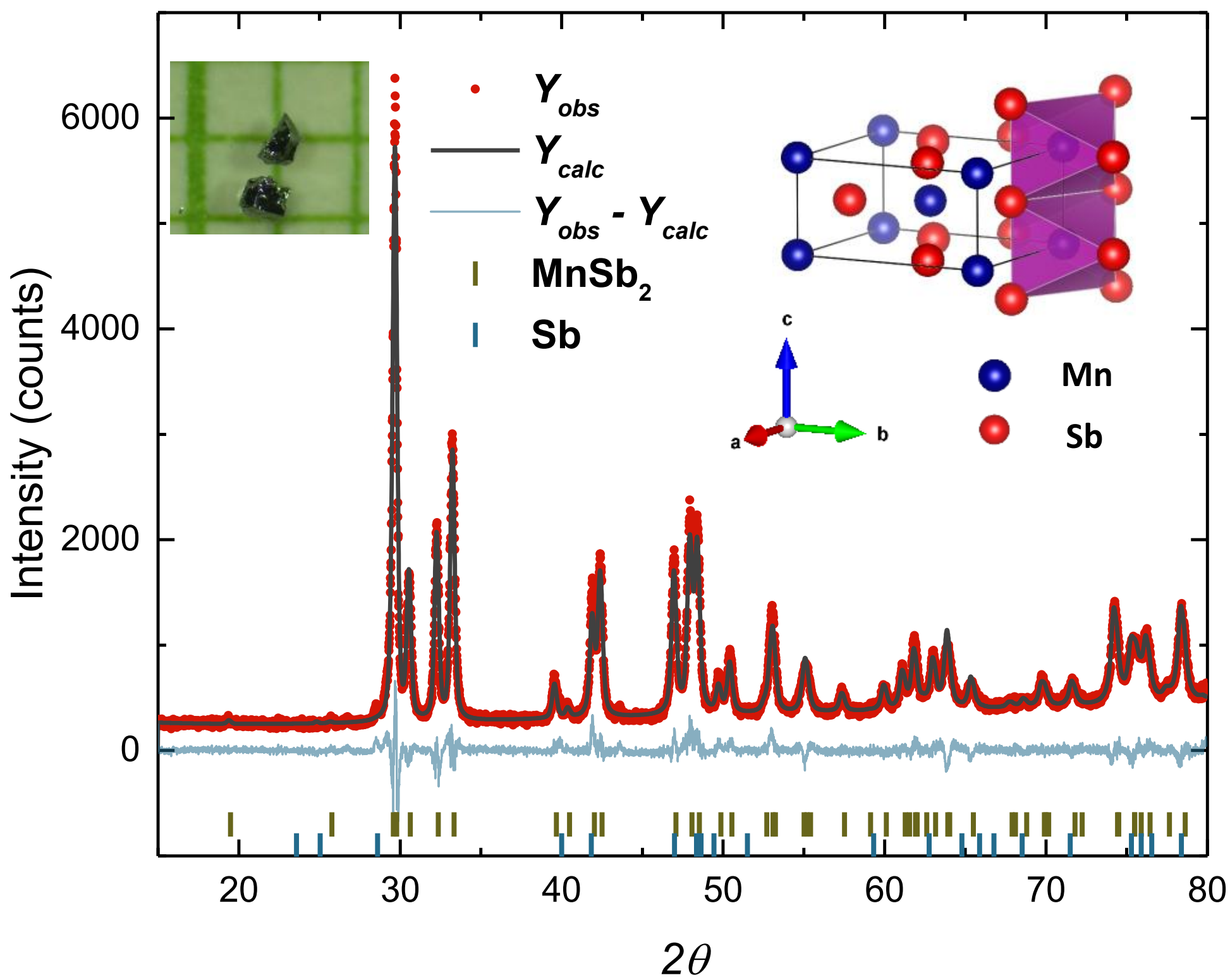

**Fig. 1 | Phase identification and structural characterization of $MnSb_2$.** Powder X-ray diffraction pattern of $MnSb_2$ with Rietveld refinement. Red circles indicate experimental data, the black line the calculated pattern, and the blue line the difference curve. The main phase is indexed to $MnSb_2$, with a minor impurity phase identified as elemental Sb (about 1% by weight). *Right inset*: crystal structure of $MnSb_2$. *Left inset*: optical image of representative single crystals; grid spacing is 1 mm.

To assess the phase purity of samples synthesized under high-pressure conditions, powder X-ray diffraction (PXRD) measurements were performed using a bulk sample with single crystals in the polycrystalline matrix, as shown in **Fig. 1**. The experimental data (red circles) were analyzed by Rietveld refinement (black line) using the GSAS-II software package[57], and the difference between the observed and calculated patterns is shown by the blue line. The minor phase of elemental Sb is observed. The PXRD results are consistent with single-crystal X-ray diffraction measurements and with the previous report.[39] The left inset of **Fig. 1** shows representative as-grown crystals, which display metallic luster and well-defined faceted morphologies, consistent with high crystallinity.

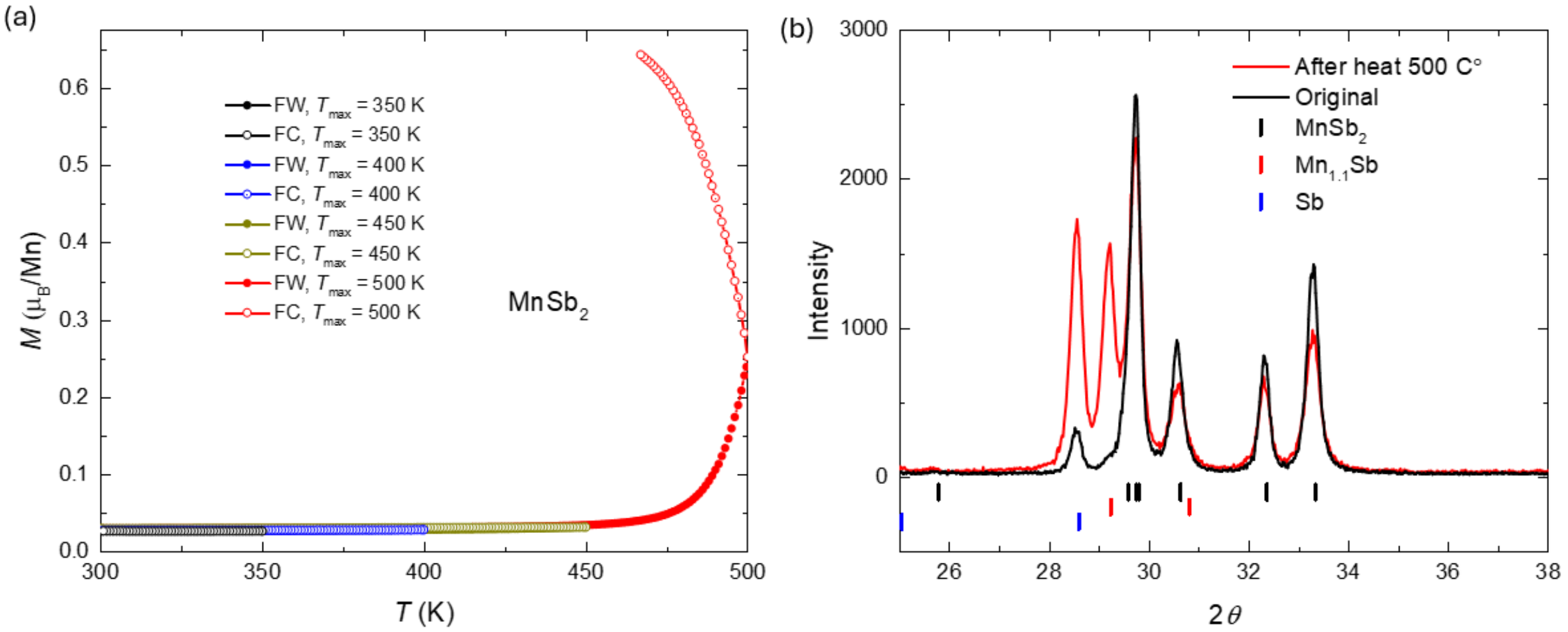

**Fig. 2 | *(a)*** Temperature-dependent magnetization of $MnSb_2$ measured under field-cooling (FC) and field-warming (FW) across different temperature ranges: 300-350 K (black), 300-400 K (blue), 300-450 K (yellow), and 300-500 K (red). A significant increase in magnetization above 450 K indicates the onset of decomposition. ***(b)*** Powder X-ray diffraction patterns collected before and after the magnetization measurements are shown in ***(a)***. The emergence of $Mn_{1.1}Sb$ reflections and an increase in Sb peak intensity after heating above 500 K confirm the thermal decomposition of $MnSb_2$.

To determine the thermal stability range of $MnSb_2$ at ambient pressure, temperature-dependent magnetization measurements were performed under both field-warming (FW) and field-cooling (FC) conditions, as shown in **Figure 2*a***. Below 450 K, the magnetization curves exhibit no significant difference between the FW and FC protocols, indicating thermal and magnetic stability within this temperature range. However, above 450 K, a pronounced increase in magnetic moment is observed, suggesting the onset of decomposition and the formation of $Mn_{1.1}Sb$, a known ferromagnetic impurity phase. This observation is corroborated by powder X-ray diffraction data collected before and after the magnetization measurements. As shown in **Figure 2*b***, additional reflections corresponding to $Mn_{1.1}Sb$ emerge, and the intensity of the Sb peak increases significantly after heating above 500 K, confirming the partial decomposition of $MnSb_2$. These results indicate that while $MnSb_2$ is metastable at ambient pressure, it remains structurally and magnetically stable at temperatures up to ~450 K, enabling long-term low-temperature studies without phase transformation. Moreover, within the temperature range of 1.8 K to 300 K, magnetization measurements (see **Figure 3**) show no signatures of ferromagnetic or other phase transitions in $MnSb_2$. The field-dependent magnetization displays predominant paramagnetic behavior with minor ferromagnetic contributions attributable to trace $Mn_{1.1}Sb$ impurities. Collectively, these results confirm that $MnSb_2$ is suitable for extended investigations at low temperatures under ambient conditions, without undergoing structural or magnetic phase transitions.

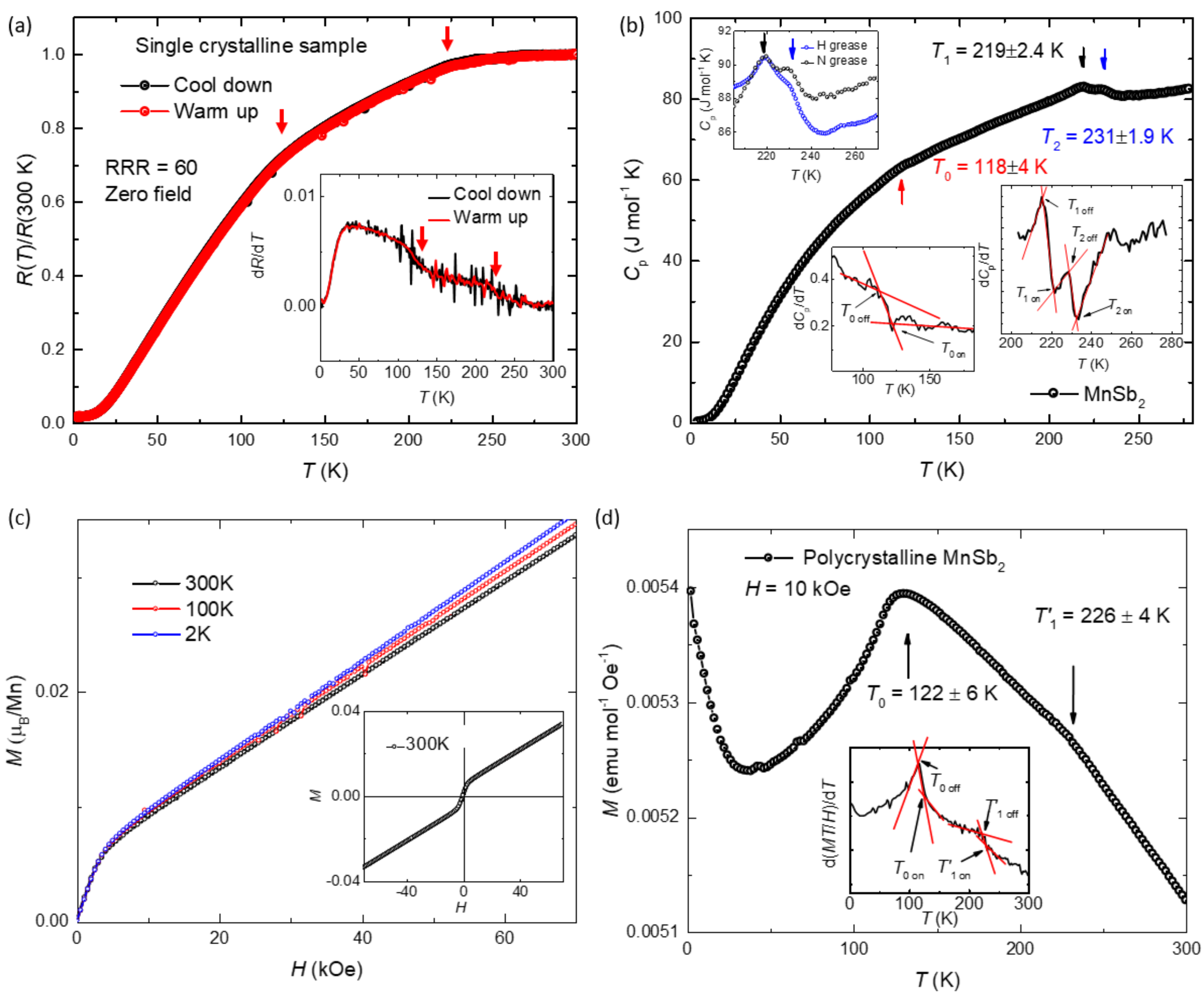

**Fig. 3 | Electrical transport, specific heat, and magnetization of $MnSb_2$. (*a*)** Electrical resistance of single-crystalline $MnSb_2$ measured upon cooling (black) and warming (red) between 300 K and 1.8 K, showing negligible thermal hysteresis. *Inset*: Temperature derivative $dR/dT$ with two anomalies. **(*b*)** Temperature-dependent specific heat showing three anomalies at $T_0$, $T_1$, and $T_2$, defined by the midpoint between onset and offset temperatures, shown in the right insets. Data from 200-280 K are obtained by point-by-point background subtraction of N-grease measurements. The temperature range is from 4K -280 K. The left inset shows a comparison of H-grease and N-grease measurements. $T_1$, and $T_2$ anomalies are clearly shown in both measurements. **(*c*)** Field-dependent magnetization (*M-H*) of polycrystalline $MnSb_2$ measured at selected temperatures. Inset: representative *M-H* curves at 300 K. **(*d*)** Temperature-dependent magnetization (*M-T*) of polycrystalline $MnSb_2$ measured under applied magnetic fields. *Inset*: temperature derivative of $M*T$, showing two anomalies between 1.8 K and 300 K. The black arrow indicates the anomalies shown in the measurement and the criteria of anomalies' temperature and width.

To probe the magnetic and electronic properties of $MnSb_2$, temperature-dependent electrical resistance measurements were performed on single-crystalline samples upon cooling from 300 K to 1.8 K and during subsequent warming, as shown in **Fig. 3*a***. The resistance curves from the cooling and warming cycles overlap closely, indicating negligible thermal hysteresis. The residual resistance ratio (RRR), defined as R(300K)/R(1.8K), is approximately 60. The inset of **Fig. 3*a*** displays the temperature derivative of the resistance ($dR/dT$). There are two clear breaks in slope visible in the raw data and two broad steps in $dR/dT$ around 125 K and 220 K, as indicated with the arrows in **Fig. 3*a*** and the inset. Complementary thermodynamic information is provided by temperature-dependent specific heat measurements on polycrystalline $MnSb_2$ performed at zero magnetic field. As shown in **Fig. 3*b***, three reproducible anomalies are observed at approximately 118 K ($T_0$), 219 K ($T_1$), and 231 K ($T_2$). These features are independently confirmed using an alternative measurement protocol over a different temperature range. Notably, the application of a magnetic field of 90 kOe produces negligible changes in specific heat (**Fig. S3*a***), demonstrating that the transition is largely insensitive to external magnetic fields. Integration of the excess specific heat yields a total magnetic entropy change of approximately $R\ln 6$ (**Fig. S3*b***) using the reference of $FeSb_2$ specific heat measurement, which is close to the Mn high-spin state. Curiously, the weak field dependence of the upper specific heat feature might suggest an antiferromagnetic transition that couples only weakly to uniform magnetic fields. To assess the uniform magnetic response of $MnSb_2$ and the presence of any net magnetic moment, magnetic susceptibility and field-dependent magnetization measurements were performed. **Fig. 3*c*** shows the field-dependent magnetization of polycrystalline $MnSb_2$ measured up to 70 kOe at selected temperatures. The magnetization exhibits only weak temperature dependence, remains unsaturated over the entire field range, and displays a very small net moment. A weak ferromagnetic contribution is observed and attributed to trace amounts (<1%) of the $Mn_{1.1}Sb$ impurity, which is below the detection limit of X-ray diffraction but was detected by NMR measurements (**Fig. S5**). The inset shows representative $M(H)$ curves measured at 300 K in five quadrants, confirming the absence of intrinsic ferromagnetic hysteresis. **Fig. 3*d*** presents the temperature-dependent magnetization of polycrystalline $MnSb_2$ measured under a 10 kOe applied field. There are two anomalies between 2 K and 300 K, $T_0$ = 122 K and $T'_1$ = 226 K, indicated by the black arrows. This observation is consistent with electric transport and the heat-capacity measurements. The inset displays the temperature derivative of $M*T$, which shows two anomalies. This behavior is consistent with an amplitude-modulated antiferromagnetic ground state and highlights the limited sensitivity of uniform magnetization measurements to this type of magnetic order as well as a relatively large contribution to $M(T)$ from a ferromagnetic impurity. To elucidate the microscopic origin of this magnetic transition, we next turn to neutron scattering measurements, which directly probe the magnetic ordering wave vector and spatial distribution of the ordered moments.

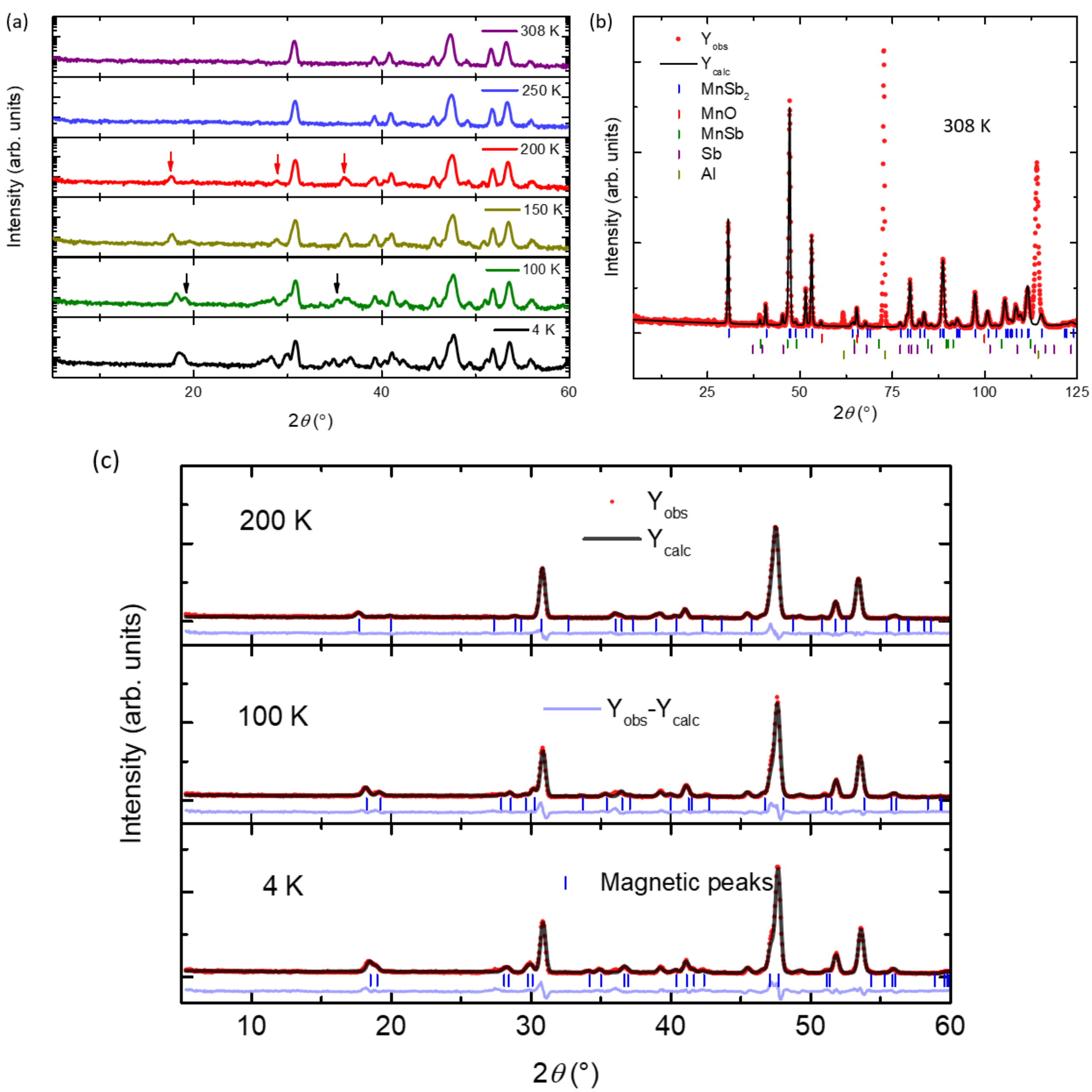

**Fig. 4 | Neutron diffraction and magnetic structure of $MnSb_2$. (*a*)** Powder neutron diffraction patterns of $MnSb_2$ collected at 308, 250, 200, 150, 100, and 4 K (logarithmic intensity scale). **(*b*, *c*)** Rietveld refinements of the powder neutron diffraction data collected at 308 K(an aluminum

can contributes the impurity peaks of Al in **Fig**. **4*b***), 200 K, 100 K, and 4 K, illustrating the emergence of magnetic reflections upon cooling.

Powder neutron diffraction measurements on $MnSb_2$ were carried out between 308 K and 4 K (**Fig. 4**). Upon cooling below ~200 K, well below the ~220 K upper magnetic transitions, additional Bragg reflections (shown in the red arrows) of magnetic origin emerge (**Fig. 4*a***). Similar magnetic peaks are found at 150 K and new peaks (shown in the black arrows) at 100K and 4 K, temperatures below the lower magnetic transition.  Refinements of diffraction patterns collected above (**Figs. 4*b***) and below (**Fig. 4*c***) the show that orthorhombic $MnSb_2$ develops a complex magnetic ground state that evolves with temperature. Two irreducible representations can be utilized to analyze the structure that is related by symmetry.

The magnetic powder refinement allows many magnetic structures to fit reasonably well. Although most structures are aligned with a spin density wave formation, trying a helical magnetic structure overall results in a lower quality of fit. The multiple models that can describe this system make it challenging to determine the direction of the spin density wave specifically and require polarized neutrons to better understand the structure. However, in many models, the Mn moment reaches a maximum of 2 $\mu_B$ and tends to be collinear with a little orientation along the *c*-axis. The propagation vector $q$ is (0, 0.3975, 0.3783) at 200 K, and the $b$ component of the propagation vector increases as temperature decreases, and the value approaches 0.5. With the changing temperature, the modulation of the spin wave changes and typically requires changing the model from high to low temperatures, further highlighting the complexity of the magnetic order within the system.

The magnetic order of orthorhombic $MnSb_2$ is more complex than a simple helical model or a uniform sinusoidal spin-density-wave description. The neutron diffraction refinements indicate an incommensurate antiferromagnetic structure with a strongly site-dependent ordered moment. Within the limitations of powder neutron diffraction, the refined structure is best described as predominantly collinear and amplitude-modulated; however, small non-collinear components cannot be excluded because the symmetry-allowed $\Gamma3$ and $\Gamma5$ representations also allow non-collinear arrangements. Therefore, we avoid assigning $MnSb_2$ to an altermagnetic class based on the present data. Further single-crystal neutron is required to test whether $MnSb_2$ exhibits symmetry-protected spin splitting or other features associated with altermagnetism.

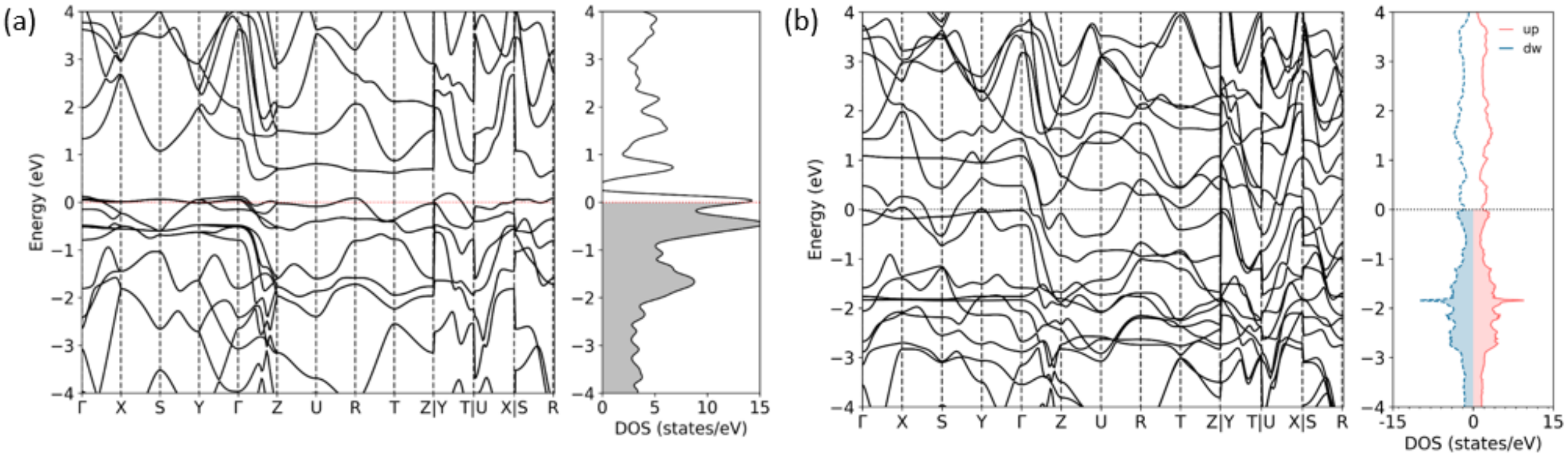

**Fig. 5 | Calculated electronic band structure of $MnSb_2$. (*a*)** Non-magnetic (spin-unpolarized) state. **(*b*)** Antiferromagnetic state.

To evaluate the magnetic ground state of $MnSb_2$, first-principles calculations were performed using the experimentally determined single-crystal structure, considering non-magnetic (NM), ferromagnetic (FM), and a simple collinear antiferromagnetic (AFM) configurations with antiparallel Mn moments in the primitive cell. Although neutron diffraction reveals a considerably more complex magnetic structure, calculations with a simple collinear AFM configuration provide a useful reference for evaluating the magnetic instability of the electronic structure. In addition, the limited experimental constraints on the full magnetic structure from unpolarized neutron diffraction and the computational cost of large magnetic supercells make such simplified calculations a practical first step. The calculated total energies (**Table S3**) show that the AFM state is the most stable, lying 80.79 meV below the NM configuration, while the FM state is also energetically favored (69.68 meV below NM), indicating close energetic competition among magnetic states. In the AFM configuration, the calculated local magnetic moment on Mn is 2.81 $\mu_B$. The calculated electronic band structures are shown in **Fig. 5**. In the NM state (**Fig. 5*a***), a flat band appears near the Fermi level, producing a pronounced peak in the density of states and signaling electronic instability. Although $MnSb_2$ shares the same nominal valence electron count as $FeSb_2$, the partial occupation of states at the Fermi level confirms metallic behavior rather than the insulating ground state of $FeSb_2$. Upon inclusion of spin polarization in the AFM state (**Fig. 5*b***), the flat band shifts below the Fermi level, and the density of states at the Fermi level is strongly reduced, leading to the formation of a pseudogap and a more stable electronic structure. Importantly, the AFM state breaks time-reversal symmetry while preserving a net zero magnetization, a symmetry condition that can allow momentum-dependent spin splitting of electronic bands in low-symmetry crystals. This symmetry condition is consistent with the collinear, finite-q antiferromagnetic order and orthogonal sublattice anisotropy resolved by neutron diffraction, although a detailed analysis of altermagnetic band splitting is beyond the scope of the present calculations. Together, these results indicate that $MnSb_2$ satisfies the essential energetic and symmetrical prerequisites for altermagnetic behavior, motivating future momentum-resolved experimental and theoretical studies.

In summary, altermagnets have recently emerged as a distinct class of magnetic materials that combine collinear antiferromagnetic order with spin-split electronic band structures and zero net magnetization, offering new opportunities for quantum and spintronic functionalities. In this work, we establish $MnSb_2$ as a chemically clean, pressure-stabilized marcasite-type compound hosting an unconventional magnetic ground state. Using high-pressure synthesis, thermodynamic measurements, neutron diffraction, and first-principles calculations, we show that $MnSb_2$ does not develop a single, rigid magnetic order below a well-defined $T_N$. Instead, magnetic entropy begins to be released near ~230 K, accompanied by the appearance of incommensurate magnetic Bragg reflections below ~ 200 K. A second thermodynamic anomaly near ~ 118 K coincides with the emergence of additional magnetic peaks, indicating a reconstruction of the ordered state. Remarkably, specific heat and neutron refinements reveal that the magnetic structure continues to evolve throughout the ordered regime: the propagation vector shifts systematically with temperature (with the *b*-component trending toward 0.5). The relating ground state is described as a complex SDW with the available data do not uniquely determine the modulation direction, and polarized neutron experiments will be required for definitive resolution. These findings establish $MnSb_2$ as a rare stoichiometric platform in which an incommensurate SDW remains highly tunable at low temperature, highlighting an unusual interplay between anisotropy, competing exchange interactions, and the underlying electronic structure.

## Acknowledgments

The work at MSU was supported by the Department of Energy under DE-SC-0023648. The work at Ames National Laboratory was supported by the U.S. Department of Energy, Office of Science, Basic Energy Sciences, Materials Sciences, and Engineering Division. Ames National Laboratory is operated for the U.S. Department of Energy by Iowa State University under contract No. DE-AC02-07CH11358. C.P. in Weiwei Xie's group is supported by NSF-DMR-2422361. A portion of this research used resources at the High Flux Isotope Reactor and Spallation Neutron Source, a DOE Office of Science User Facility operated by the Oak Ridge National Laboratory.

The beam time was allocated to HB-2A on proposal number IPTS-34171.1 and POWGEN on proposal number IPTS-?????.1. A portion of this research used resources at the Spallation Neutron Source, a DOE Office of Science User Facility operated by the Oak Ridge National Laboratory. This material is based upon work supported by the U.S. Department of Energy, Office of Science, Office of Workforce Development for Teachers and Scientists, Office of Science Graduate Student Research (SCGSR) program. The SCGSR program is administered by the Oak Ridge Institute for Science and Education (ORISE) for the DOE. ORISE is managed by ORAU under contract number DESC0014664. All opinions expressed in this paper are the author's and do not necessarily reflect the policies and views of DOE, ORAU, or ORISE. No animals were harmed in the course of this research.

## Conflicts of interest

The authors declare no competing financial interest.

## TOF

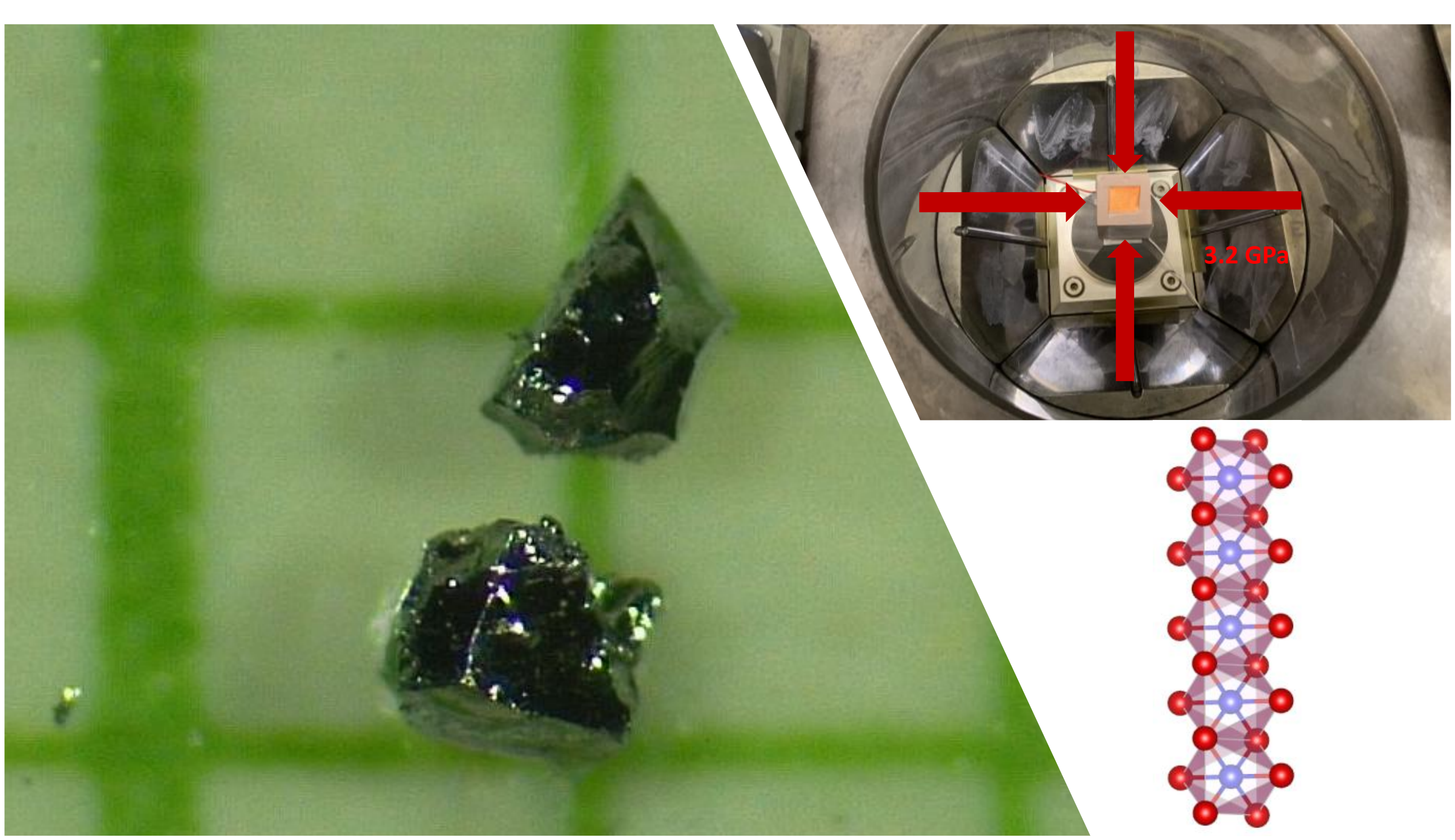

## Support Information

# Pressure-Stabilized $MnSb_2$ with Complex Incommensurate Magnetic Order

Mingyu Xu[1,2,3], Matt Boswell[1,4], Qing-Ping Ding[2], Peng Cheng[1], Aashish Sapkota[2,3], Qiang Zhang[4], Danielle Yahne[4], Sergey. L. Bud'ko[2, 3], Yuji Furukawa[2,3], Paul. C. Canfield[2,3], Raquel A. Ribeiro[2,3], Weiwei Xie[1*]

[1]Department of Chemistry, Michigan State University, East Lansing, Michigan 48824, USA
[2]Ames National Laboratory, Iowa State University, Ames, Iowa 50011, USA
[3]Department of Physics and Astronomy, Iowa State University, Ames, Iowa 50011, USA
[4]Neutron Scattering Division, Oak Ridge National Laboratory, Oak Ridge, Tennessee 37831, USA

**Table S1.** The crystal structure and refinement of $MnSb_2$ at room temperature. Values in parentheses are estimated standard deviation from refinement.

| **Chemical Formula** | **$MnSb_2$** |
|---|---|
| Formula Weight | 298.44 g/mol |
| Space Group | *Pnnm* |
| lUnit Cell dimensions | $a$ = 6.0227(3) Å<br>$b$ = 6.8893(3) Å<br>$c$ = 3.3240(2) Å |
| Volume | 137.918(12) $Å^3$ |
| Density (calculated) | 7.186 $g/cm^3$ |
| Absorption coefficient | 23.579 $mm^{-1}$ |
| F (000) | 254 |
| 2θ range | 9.00 to 80.26° |
| Reflections collected | 3540 |
| Independent reflections | 478 [$R_{int}$ = 0.0561] |
| Refinement method | Full-matrix least-squares on $F^2$ |
| Data/restraints/parameters | 478/0/11 |
| Final $R$ indices | $R_1$ (I>2σ(I)) = 0.0282; $wR_2$ (I > 2 σ(I)) = 0.0795<br>$R_1$ (all) = 0.0289; $wR_2$ (all) = 0.0799 |
| Largest diff. peak and hole | +2.072 $e^-/Å^3$ and -1.637 $e^-/Å^3$ |
| R. M. S. deviation from mean | 0.428 $e^-/Å^3$ |
| Goodness-of-fit on $F^2$ | 1.229 |

**Table S2.** Atomic coordinates and equivalent isotropic atomic displacement parameters ($Å^2$) $MnSb_2$. ($U_{eq}$ is defined as one-third of the trace of the orthogonalized $U_{ij}$ tensor.) Values in parentheses are estimated standard deviations from refinement.

| **Atoms** | **Wyck.** | ***x*** | ***y*** | ***z*** | **Occ.** | **$U_{eq}$** |
|---|---|---|---|---|---|---|
| **Sb** | 4*g* | 0.17921(5) | 0.36252(4) | 0 | 1 | 0.01228(11) |
| **Mn** | 2*a* | 0 | 0 | 0 | 1 | 0.01182(17) |

**Table S3**. Calculated the total energy of different magnetic models.

| | **NM** | **FM** | **AFM** |
|---|---|---|---|
| Magnetic Model | 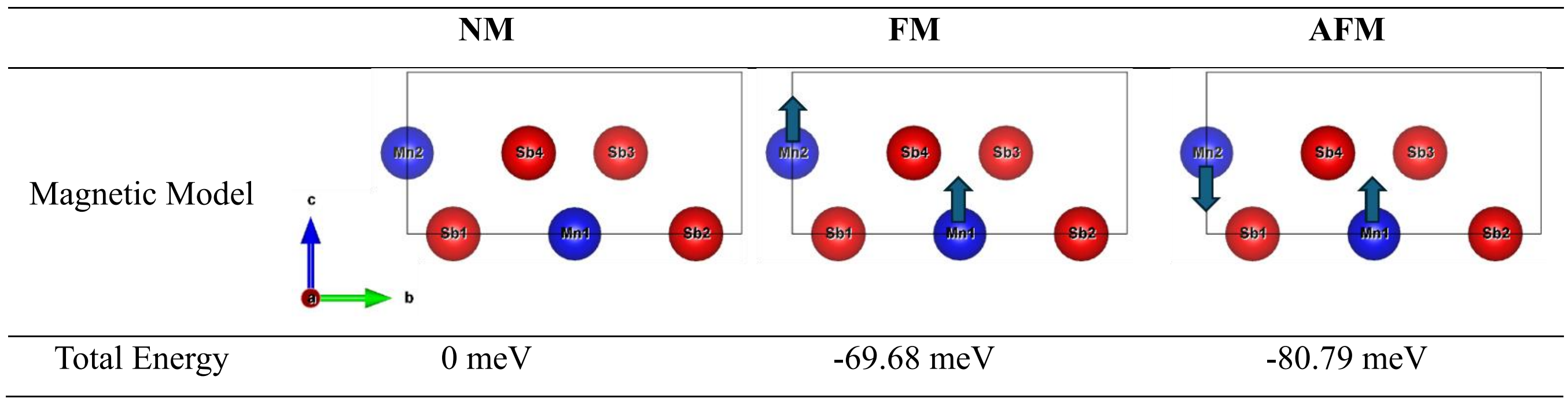 | | |
| Total Energy | 0 meV | -69.68 meV | -80.79 meV |

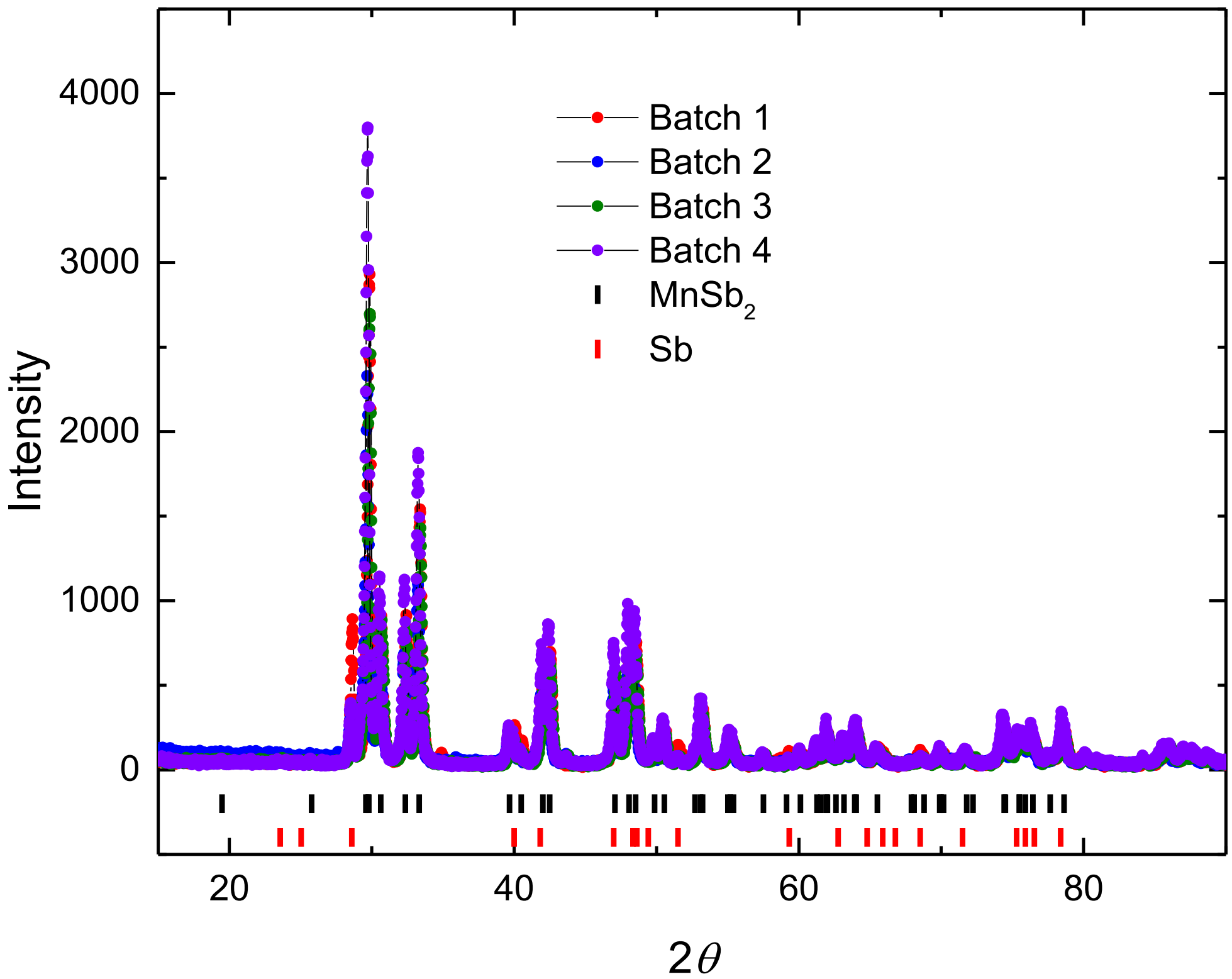

**Figure S1**, Powder XRD results from the different batches. The sample for XRD is a part of the chunk directly from synthesis without the separation of single crystals.

**Figure S1** shows the powder XRD results from the different batches. The sample for XRD is a part of the chunk directly from synthesis without the separation of single crystals. To complete the reaction of Mn, a small excess of Sb (~1 wt%) was added.

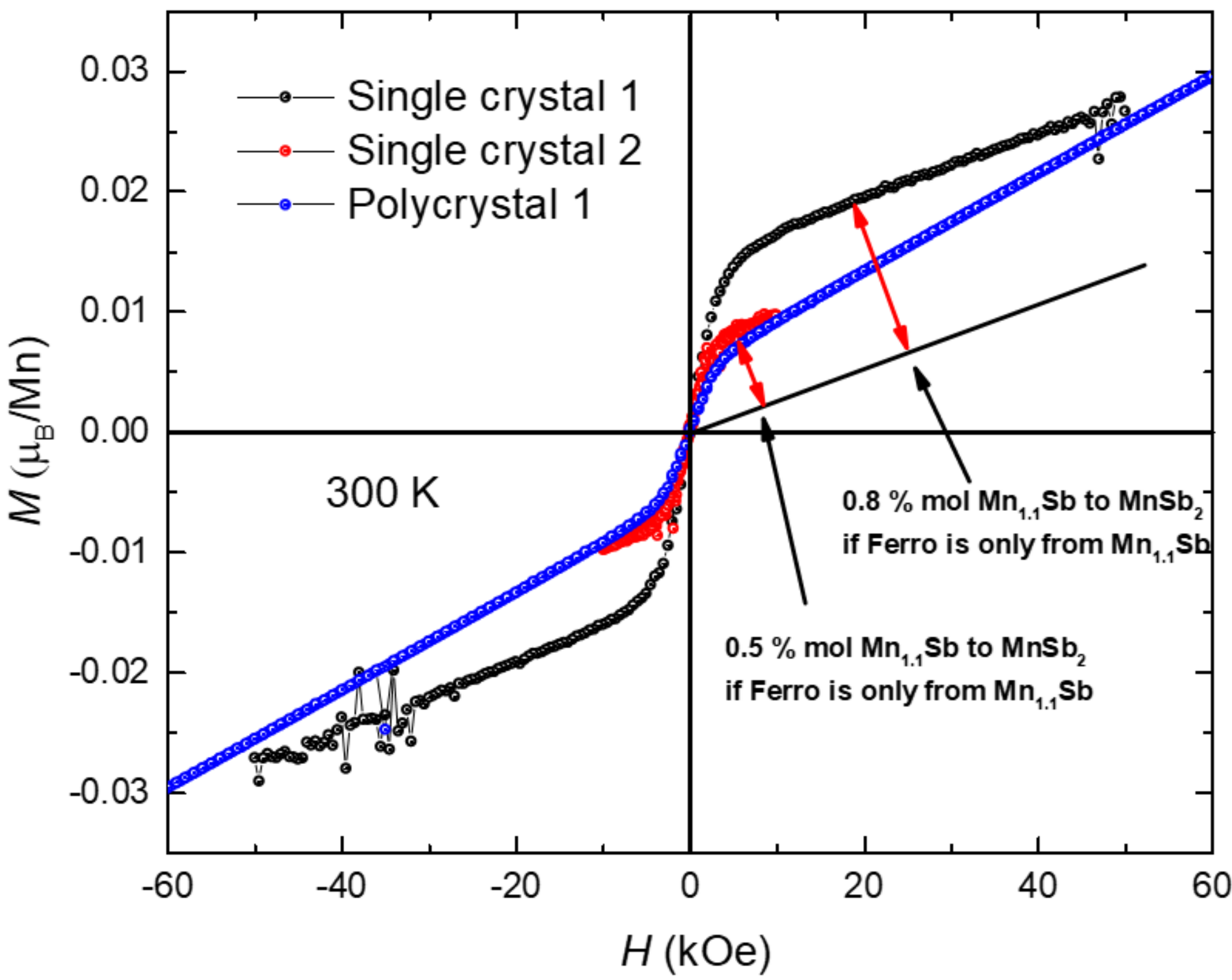

**Figure S2**, M(H) data of single crystals and polycrystals at 300 K. The mole ratio of $Mn_{1.1}Sb$ is estimated by assuming all the ferromagnetic single crystals come from $Mn_{1.1}Sb$.

**Figure S2** shows *M* (*H*) data of single crystals and polycrystals at 300 K. The mole ratio of Mn1.1Sb is estimated by assuming all the ferromagnetic single crystals come from MnSb.

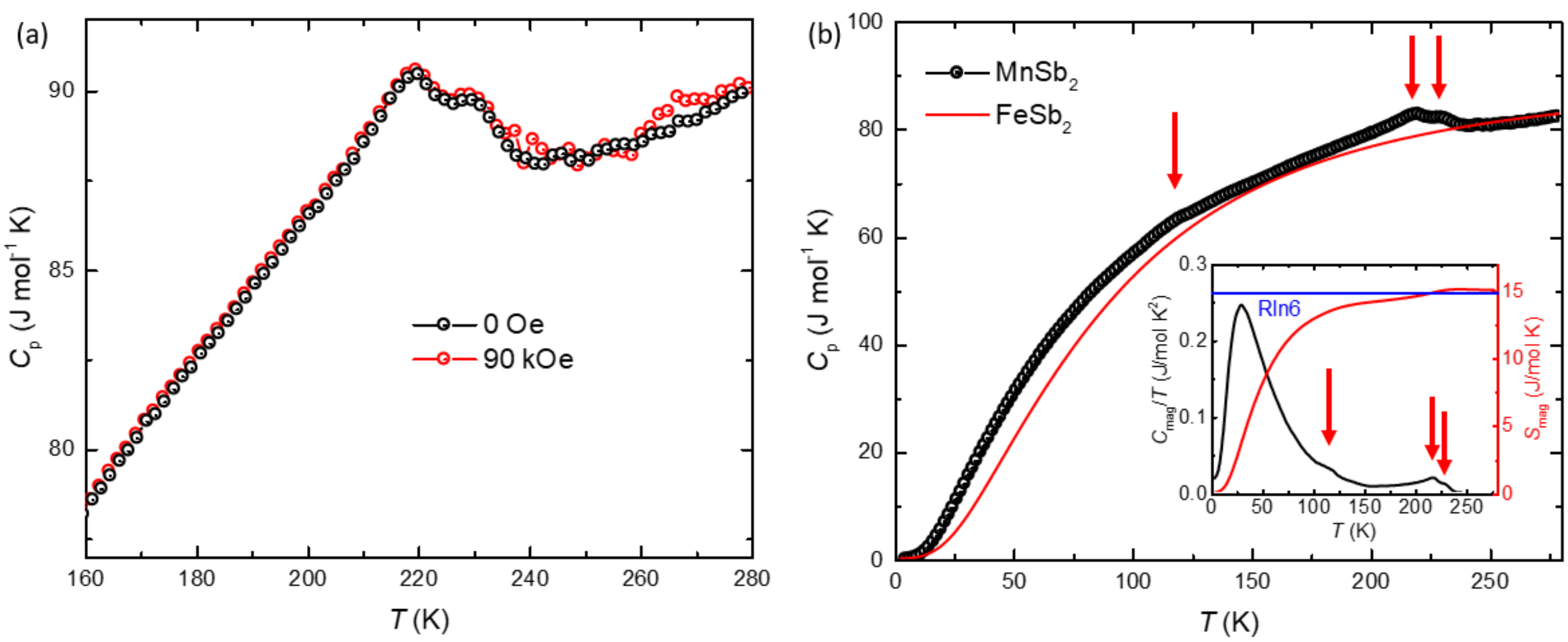

**Figure S3**, Temperature-dependent specific heat was measured in the temperature range from 150 K to 280 K and 2 K to 200 K. The measurement is taken on the N-grease with addenda taken with the same data points as the sample measurements in the temperature range from 150 K to 280 K. There is no clear difference between the field at 0 Oe or 90 kOe. Magnetic entropy is calculated with reference to $FeSb_2$ specific heat measurement.

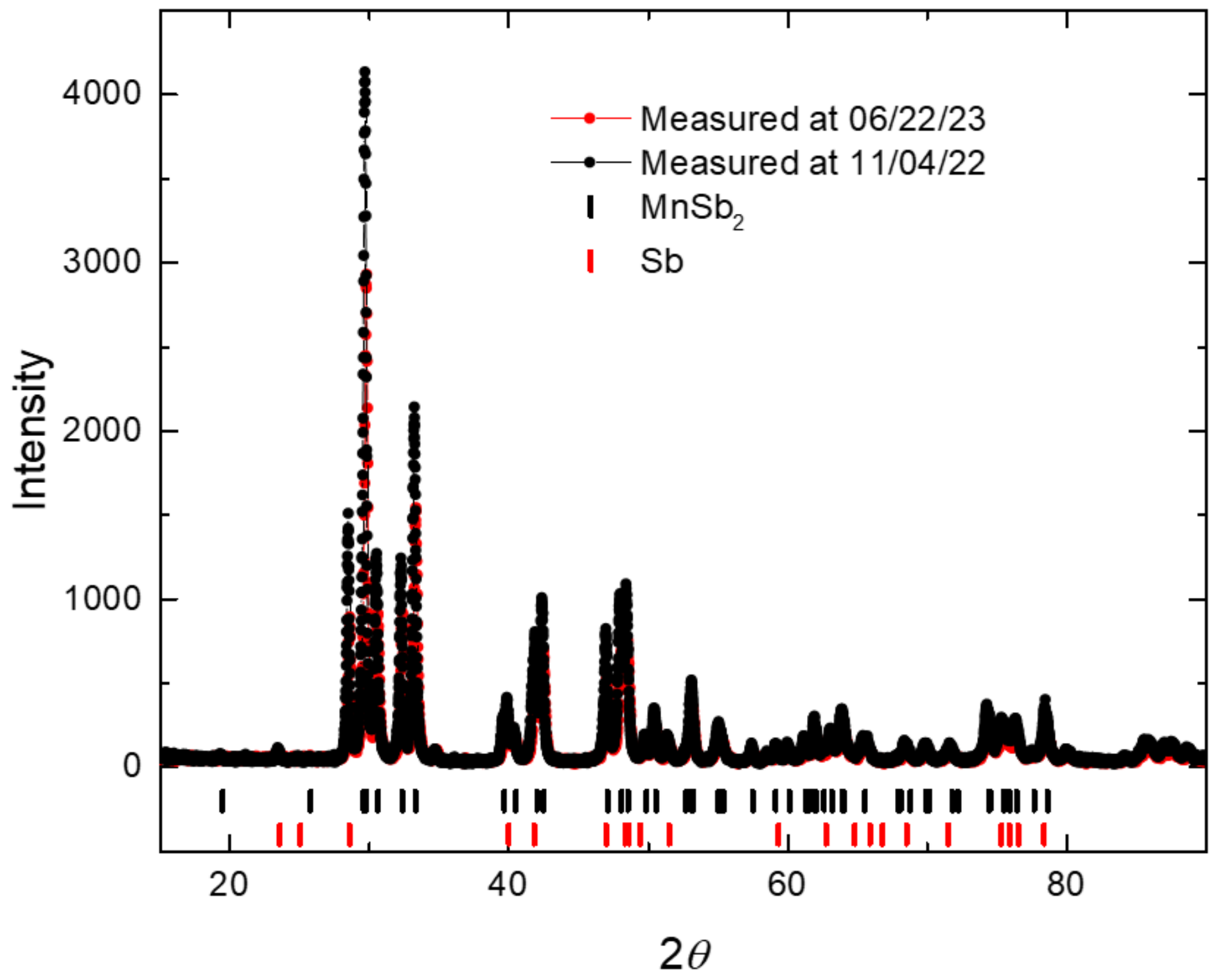

**Figure S4.** Powder XRD results from the same batch. The interval of the time is around six months. The highest intensity peak normalizes the intensity.

To assess the ambient stability of $MnSb_2$ over time, **Figure S4** shows the normalized powder X-ray diffraction (XRD) patterns of the same sample batch measured immediately after synthesis (black) and after six months of air exposure at room temperature (red). Despite $MnSb_2$ being a metastable phase at ambient pressure, no discernible changes in the diffraction pattern are observed, indicating that the phase remains structurally stable under ambient conditions for 6 months. This result suggests that $MnSb_2$ is stable at or below room temperature, with no evidence of decomposition or transformation over six months.

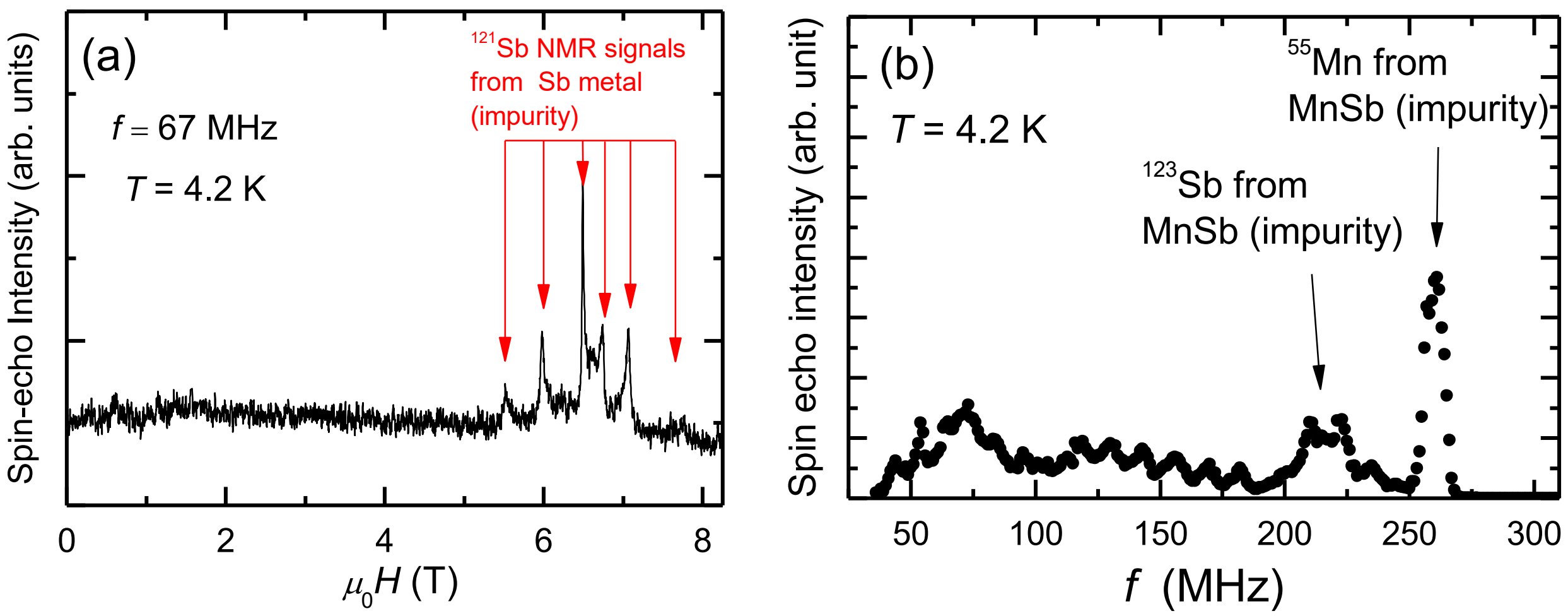

**Figure S5**. (a) NMR spectrum measured at $T$ = 4.2 K by sweeping the magnetic field on polycrystalline powder samples. (b) Zero-field NMR spectrum at $T$ = 4.2 K on polycrystalline powder samples.

We attempted to get information about the magnetic ground state of $MnSb_2$ from $^{55}$Mn, $^{121}$Sb, and $^{123}$Sb NMR measurements at a low temperature of $T$ = 4.2 K using the same sample from neutron diffraction measurements. **Figure S5*a*** shows the NMR spectrum, where a very broad spectrum from the highest magnetic field of our magnet (8.25 T) to zero field was observed. The relatively sharp lines detected from 5.5 T to 7.5 T (indicated by the red arrows) were assigned to a typical quadrupolar split $^{121}$Sb NMR power lines from the impurity of Sb metal[1]. The intrinsic broad spectrum indicates that the hyperfine fields at the Mn and Sb sites are widely distributed in the magnetically ordered state, consistent with the inhomogeneous magnetic state as determined by the neutron diffraction measurements. Such an inhomogeneous magnetic state is also suggested by the very broad zero-field NMR spectrum shown in **Figure S5*b***. The signals around 250-260 MHz and 200-230 MHz are from $^{55}$Mn and $^{123}$Sb, respectively, in the impurity of the ferromagnetic $Mn_{1.1}Sb$[2]. Although the broad spectrum is not fully resolved at present, we consider that the signals below 200 MHz are due to $^{55}$Mn and $^{121}$Sb NMR signals in antiferromagnetically ordered $MnSb_2$. This again suggests the large distributions of hyperfine fields at the Mn and Sb sites, consistent with the NMR spectrum, indicating the inhomogeneous magnetic state.

To investigate the possibility that high-pressure synthesized $MnSb_2$ exhibits magnetic ordering above room temperature ($T_n$ > 300 K), powder neutron diffraction measurements were conducted to detect either additional magnetic Bragg reflections or intensity enhancements of the nuclear Bragg peaks consistent with the P*nmm* crystal symmetry. In general, antiferromagnetic (AFM) ordering gives rise to new magnetic reflections, while ferromagnetic ordering manifests as enhanced intensity at the positions of nuclear Bragg peaks. **Figure S6*a*** presents the neutron diffraction pattern collected at 300 K, along with the corresponding Rietveld refinement. All observed peaks are consistent with the P*nmm* symmetry of $MnSb_2$, and no additional magnetic peaks are detected. The refined lattice parameters from the neutron diffraction data are $a$ = 6.0162(2) Å, $b$ = 6.8824(3) Å, and $c$ = 3.3226(1) Å, which are in good agreement with the values obtained from the X-ray diffraction refinement. Minor impurity phases were identified as MnSb (2.7 wt%), MnO (0.3 wt%), Sb (3.6 wt%), and boron nitride (BN, 0.07 wt%), which account for the weak reflections not indexed to the primary phase. Statistical parameters from the refinement, including all the phases in the model, are $R_{wp}$ = 6.0% and GOF= 3.2, indicating a reasonable refinement. The site occupancy, atomic coordinates, and isotropic displacement parameters obtained from the refinement are summarized in **Table S4**.

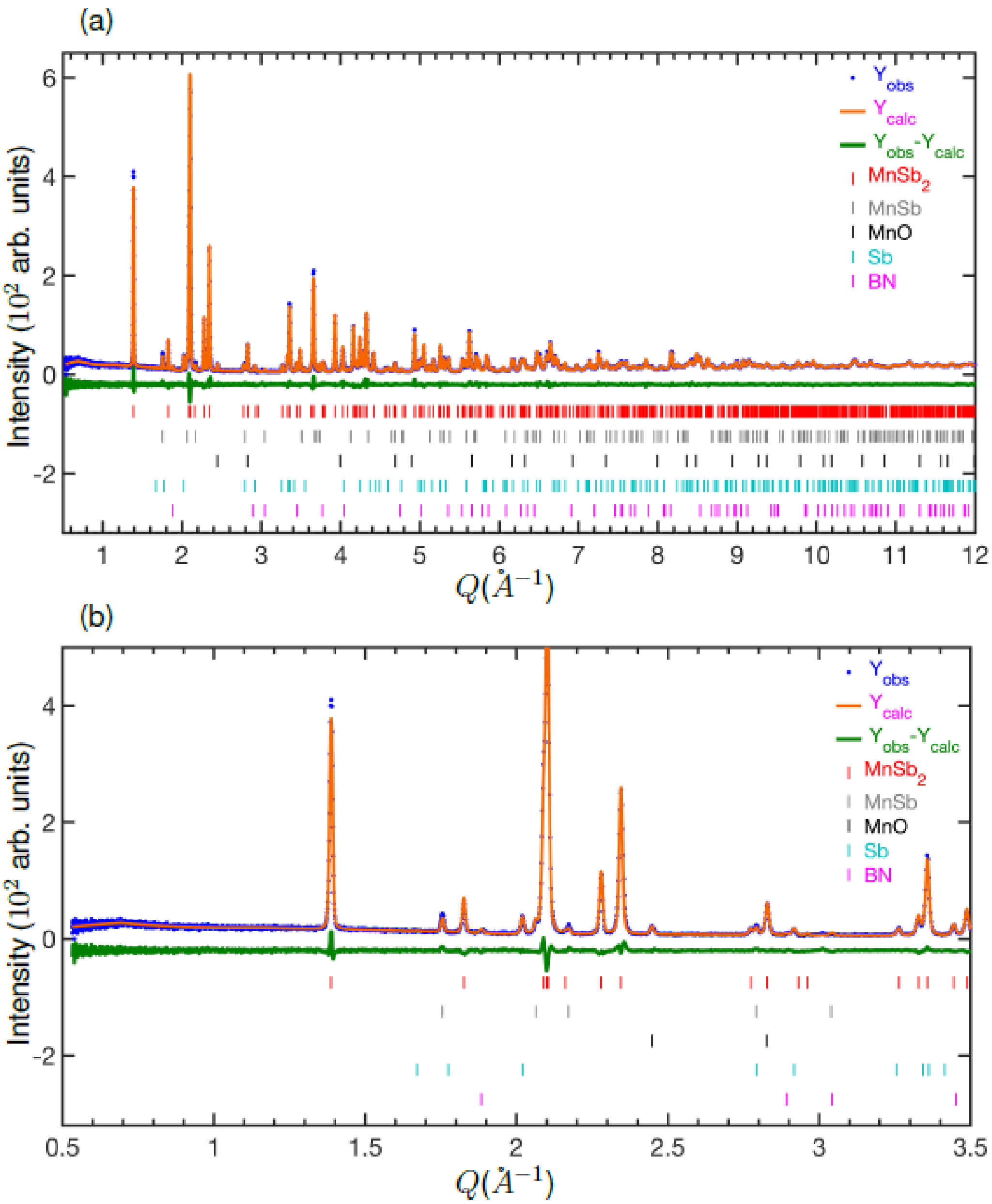

**Figure S6.** ***(a)*** Rietveld refinement of the 300 K powder neutron diffraction data for $MnSb_2$, yielding a weighted profile R-factor ($R_{wp}$) of 6.0% and a goodness of fit (GOF) of 3.2. The blue dots represent the experimental data, while the orange line corresponds to the calculated pattern. The green line indicates the difference between the observed and calculated intensities. Vertical tick marks represent the allowed Bragg reflection positions for the identified phases: $MnSb_2$ (red), Sb (gray), MnSb (black), MnO (cyan), and boron nitride (BN, magenta). ***(b)*** Enlarged view of the low-Q region from panel ***(a)***, emphasizing the absence of additional magnetic Bragg peaks, indicating no long-range magnetic ordering at room temperature.

**Table S4.** Refined atomic coordinates and the equivalent isotropic displacement parameters of $MnSb_2$ at 300 K.

| **Atoms** | **Wyck.** | ***x*** | ***y*** | ***z*** | **Occ.** | $U_{eq}$ |
|---|---|---|---|---|---|---|
| **Sb** | 4*g* | 0.1799(1) | 0.3624(1) | 0 | 1 | 0.0080(2) |
| **Mn** | 2*a* | 0 | 0 | 0 | 1 | 0.0060(3) |

**Table S5.** Basis vectors for the space group *Pnnm* with k = (0, 0, 0). The decomposition of the magnetic representation for the Mn site (0, 0, 0) is $\Gamma_{Mag} = 1\Gamma_1^1 + 2\Gamma_3^1 + 2\Gamma_5^1 + 1\Gamma_7^1$. The atoms of the nonprimitive basis are defined according to 1: (0, 0, 0), 2: (1/2, 1/2, 1/2).

| IR | BV | Atom | BV components | | | | | |
|---|---|---|---|---|---|---|---|---|
| | | | $m_{\parallel a}$ | $m_{\parallel b}$ | $m_{\parallel c}$ | $im_{\parallel a}$ | $im_{\parallel b}$ | $im_{\parallel c}$ |
| $\Gamma_1$ | $\psi_1$ | 1 | 0 | 0 | 4 | 0 | 0 | 0 |
| | | 2 | 0 | 0 | -4 | 0 | 0 | 0 |
| $\Gamma_3$ | $\psi_2$ | 1 | 4 | 0 | 0 | 0 | 0 | 0 |
| | | 2 | 4 | 0 | 0 | 0 | 0 | 0 |
| | $\psi_3$ | 1 | 0 | 4 | 0 | 0 | 0 | 0 |
| | | 2 | 0 | -4 | 0 | 0 | 0 | 0 |
| $\Gamma_5$ | $\psi_4$ | 1 | 4 | 0 | 0 | 0 | 0 | 0 |
| | | 2 | -4 | 0 | 0 | 0 | 0 | 0 |
| | $\psi_5$ | 1 | 0 | 4 | 0 | 0 | 0 | 0 |
| | | 2 | 0 | 4 | 0 | 0 | 0 | 0 |
| $\Gamma_7$ | $\psi_6$ | 1 | 0 | 0 | 4 | 0 | 0 | 0 |
| | | 2 | 0 | 0 | 4 | 0 | 0 | 0 |

**Table S6.** Symmetry-allowed magnetic basis vectors obtained from representation analysis for the Mn1 and Mn2 sublattices, obtained from Rietveld refinements of the powder neutron diffraction data collected at 200 K. For each sublattice, the basis vectors $\psi 1$, $\psi 2$, and $\psi 3$ correspond to magnetic moment components along the crystallographic $a$, $b$, and $c$ axes, respectively.

| Mn1 | *a* | *b* | *c* |
|---|---|---|---|
| $\psi 1$ | 1 | 0 | 0 |
| $\psi 2$ | 0 | 1 | 0 |
| $\psi 3$ | 0 | 0 | 1 |
| Mn2 | | | |
| $\psi 1$ | 1 | 0 | 0 |
| $\psi 2$ | 0 | 1 | 0 |
| $\psi 3$ | 0 | 0 | 1 |

To further probe potential magnetic contributions in $MnSb_2$, **Figure S6*b*** highlights the low-$Q$ region of the neutron diffraction pattern, where magnetic scattering is typically enhanced due to the $Q$-dependence of the magnetic form factor. The absence of additional Bragg reflections in this region provides compelling evidence against the presence of long-range magnetic ordering at room temperature. In particular, the lack of magnetic peaks strongly suggests that antiferromagnetic (AFM) ordering with a non-zero propagation vector ($k \neq (0, 0, 0)$) is absent in this compound. Nonetheless, the possibility of ferromagnetic ordering corresponding to $k = (0, 0, 0)$ cannot be excluded solely based on the absence of additional reflections. In such a case, magnetic ordering would manifest as an enhancement of nuclear Bragg peak intensities. However, our Rietveld refinement shows no observable intensity increase at the nuclear peak positions, and no systematic deviation between observed and calculated intensities was detected. To definitively assess the presence of ferromagnetic ordering and to estimate an upper bound for the ordered Mn moment, magnetic structure refinements were performed using the FullProf suite with magnetic subgroups of the *Pnmm* space group. Symmetry-allowed magnetic configurations were derived from a group-subgroup analysis of the propagation vector $k = (0, 0, 0)$, implemented via SARAh-Representational Analysis. Four irreducible representations (IRs) and their corresponding basis vectors (BVs) were identified, as listed in **Table S5**. Among them, IRs $\Gamma_3$ (with BV $\psi_2$) and $\Gamma_5$ (with BV $\psi_5$), which describe ferromagnetic alignment of Mn moments within the *ab*-plane, produced the best magnetic refinements, with magnetic $R$-factors ($R_{mag}$) below 5. Despite these fits, the refined ordered moment was consistently small, with a total magnetic moment of $\mu_{tot}$ that should be much smaller than 0.2 $\mu_B$/Mn, indicating the absence of significant static magnetic ordering in $MnSb_2$ at room temperature.

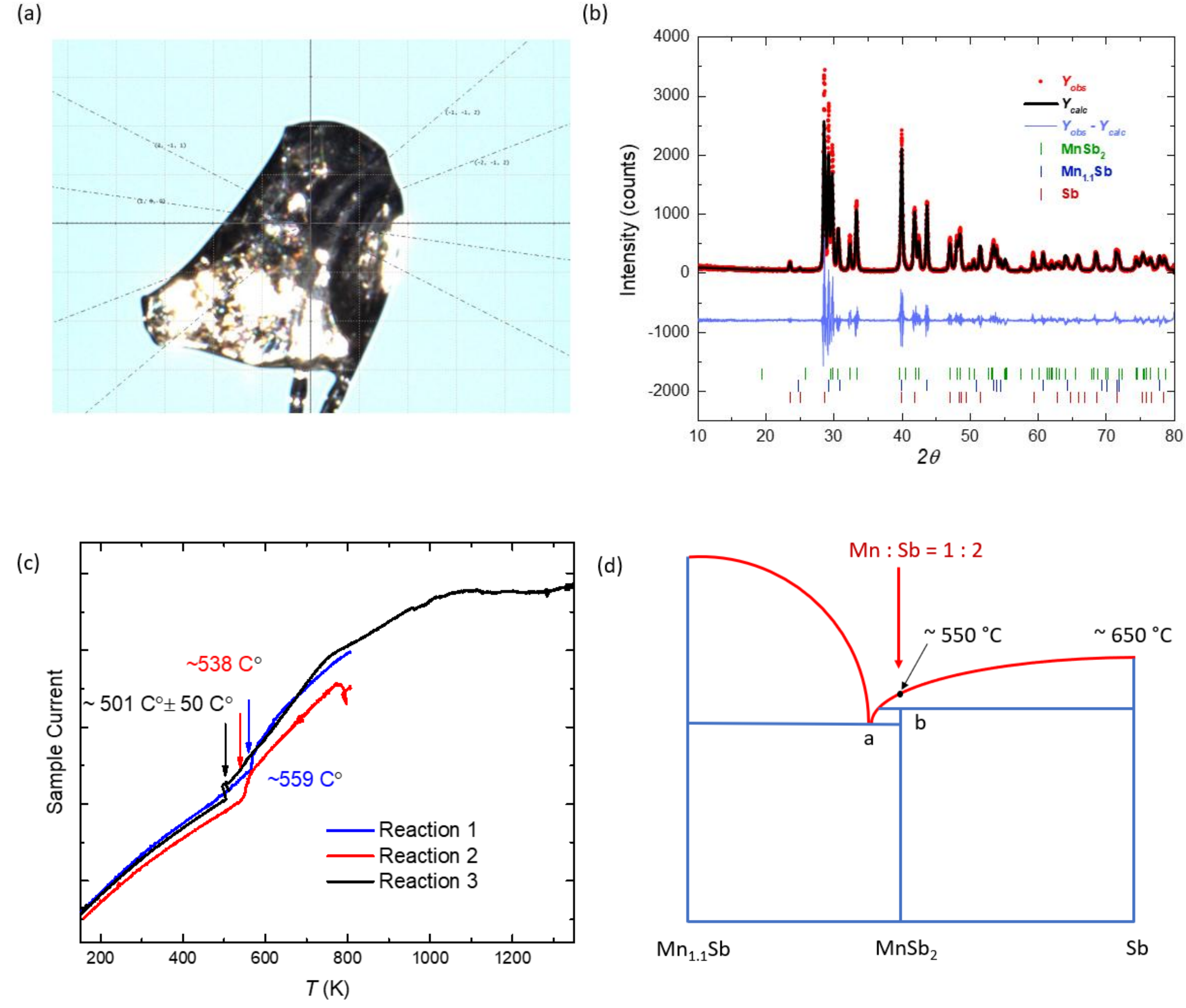

**Figure S7.** **(*a*)** The sample used to determine the crystallographic direction of $MnSb_2$ is shown. **(*b*)** The PXRD data are shown for solid-state reactions at temperatures above 700 °C. **(*c*)** The sample current as a function of temperature is shown for different $MnSb_2$ reactions. **(*d*)** The phase diagram is shown. A point is the eutectic point, and the b point is the peritectic point. The red curve indicates the liquidus lines.